\documentclass[10pt,aps,pra,amsmath,amssymb,twocolumn,superscriptaddress]{revtex4-2}

\usepackage[english]{babel}
\usepackage{CJKutf8}
\usepackage{graphicx}
\usepackage{siunitx}
\usepackage[letterpaper,total={7in,9.5in},top=0.75in,left=0.75in]{geometry}
\usepackage[colorlinks=true,linkcolor=blue,citecolor=blue,urlcolor=blue,filecolor=blue]{hyperref}
\usepackage{tikz}
\usepackage{xcolor}
\usepackage{braket}

\definecolor{lime}{HTML}{A6CE39}
\DeclareRobustCommand{\orcidicon}{
	\begin{tikzpicture}
	\draw[lime, fill=lime] (0,0) 
	circle [radius=0.16] 
	node[white] {{\fontfamily{qag}\selectfont \tiny ID}};
	\draw[white, fill=white] (-0.0625,0.095) 
	circle [radius=0.007];
	\end{tikzpicture}
	\hspace{-0.3mm}
}

\foreach \x in {A, ..., Z}{\expandafter\xdef\csname orcid\x\endcsname{\noexpand\href{https://orcid.org/\csname orcidauthor\x\endcsname}
			{\noexpand\orcidicon}}
}

\newcommand{\Sr}[1]{$^{#1}$Sr} 
\newcommand{\RedMotTransition}{${^1\mathrm{S}_0} - {^3\mathrm{P}_1}$}
\newcommand{\RedMotLinewidth}{$\SI{7.5}{\kilo \hertz}$}
\newcommand{\PumpingTransition}{${^1\mathrm{S}_0}~{\mathrm{F} = 9/2} - {^3\mathrm{P}_1}~{\mathrm{F'} = 9/2}$}
\newcommand{\CoolingTransition}{${^1\mathrm{S}_0}~{\mathrm{F} = 9/2} - {^3\mathrm{P}_1}~{\mathrm{F'} = 11/2}$}
\newcommand{\clocktransition}{${^1\mathrm{S}_0} - {^3\mathrm{P}_0}$}
\newcommand{\BlueMotTransition}{${^1\mathrm{S}_0} - {^1\mathrm{P}_1}$}
\newcommand{\BlueMotLinewidth}{$\SI{30}{\mega \hertz}$}
\newcommand{\mf}{$m_{\mathrm{F}}$}

\newcommand{\groundState}{${^1\mathrm{S}_0} \, {\mathrm{F} = 9/2}$}
\newcommand{\StirringState}{${^3\mathrm{P}_1} \, {\mathrm{F^{\prime}} = 9/2}$}

\newcommand{\GroundStateGyromagneticRatio}{$2 \pi \times \SI{183(7)}{\hertz \per \gauss}$} 

\newcommand{\fieldGradients}{$\{-0.55, 0.32, 0.23\} \, \si{\gauss \per \centi \meter}$} 
\newcommand{\MOTAtomNumber}[1][atoms]{$\SI{8.41(4) e7}{#1}$} 
\newcommand{\MOTAtomNumberAbstrac}[1][atoms]{$\SI{8.4 e7}{#1}$} 

\newcommand{\MOTLoadingTime}[1][\second]{$\SI{6.4(1)}{#1}$}
\newcommand{\MOTflux}[1][atoms \per \second]{$\SI{1.31(2)e7}{#1}$}
\newcommand{\MOTfluxAbstract}[1][atoms \per \second]{$\SI{1.3e7}{#1}$}
\newcommand{\MOTLifetime}[1][\second]{$\SI{4.84(3)}{#1}$} 
\newcommand{\MOTZtemp}[1][\micro \kelvin]{$\SI{14.2(2)}{#1}$} 
\newcommand{\MOTYtemp}[1][\micro \kelvin]{$\SI{7.5(2)}{#1}$} 
\newcommand{\MOTAvgtemp}[1][\micro \kelvin]{$\SI{12.0(1)}{#1}$} 
\newcommand{\MOTAvgtempAbstract}[1][\micro \kelvin]{$\SI{12}{#1}$} 
\newcommand{\MOTDensity}[1][^{87}Sr \per\centi\meter\cubic]{$\SI{2.12(4) e9}{#1}$}
\newcommand{\MotWidhty}[1][\milli\meter]{$\SI{0.689(8)}{#1}$}
\newcommand{\MotWidhtz}[1][\milli\meter]{$\SI{ 1.91(2)}{#1}$} 
\newcommand{\StirringvsnSti}{$2.2(3)$}

\newcommand{\BosonicMOTAtomNumber}[1][atoms]{$\SI{100.8(9)e7}{#1}$} 
\newcommand{\BosonicMOTlifetime}[1][\second]{$\SI{4.6(1)}{#1}$}
\newcommand{\BosonicMOTLoadingTime}[1][\second]{$\SI{4.4(1)}{#1}$}
\newcommand{\BosonicMOTflux}[1][atoms \per \second]{$\SI{2.29 (6) e8 }{#1}$}

\newcommand{\BosonNssoverfermionNss}{$17$}

\newcommand{\BosonicMOTZtemp}[1][\micro \kelvin]{$\SI{12.6(1)}{#1} $} 
\newcommand{\BosonicMOTYtemp}[1][\micro \kelvin]{$\SI{9(1)}{#1}$} 
\newcommand{\BosonicMOTAvgtemp}[1][\micro \kelvin]{$\SI{11.4(4)}{#1}$} 
 
\newcommand{\BosonicMOTDensity}[1][^{88}Sr \, atoms \per\milli\meter\cubic]{$\SI{1.28(2) e10}{#1}$}
\newcommand{\BosonicMOTWidhtz}[1][\milli \meter]{$\SI{2.82 (2)}{#1} $} 
\newcommand{\BosonicMOTWidhty}[1][\milli \meter]{$\SI{0.632 (8)}{#1}$}

\newcommand{\PowerHighPowerBeam}{$\SI{15.2}{\milli \watt}$}
\newcommand{\PowerLowPowerBeam}{$\SI{14.2}{\milli \watt}$}

\newcommand{\SweepTime}{$\SI{71}{\micro \second} $}
\newcommand{\OmegaSWAPAverage}{$2 \pi \times \SI{90}{\kilo \hertz}$} 

\DeclareSIUnit\gauss{G}
\DeclareSIUnit\mfu{mf}

\begin{document}

\title{A steady-state magneto-optical trap of fermionic strontium on a narrow-line transition}

\author{Rodrigo Gonz\'{a}lez Escudero$\orcidA{}$}
\author{Chun-Chia Chen (陳俊嘉)$\orcidB{}$}
\author{Shayne Bennetts$\orcidC{}$}
\author{Benjamin Pasquiou$\orcidD{}$}
\affiliation{Van der Waals-Zeeman Institute, Institute of Physics, University of Amsterdam, Science Park 904, 1098XH Amsterdam, The Netherlands}
\author{Florian Schreck$\orcidE{}$}
\email[]{FermiMOT@strontiumBEC.com}
\affiliation{Van der Waals-Zeeman Institute, Institute of Physics, University of Amsterdam, Science Park 904, 1098XH Amsterdam, The Netherlands}
\affiliation{QuSoft, Science Park 123, 1098XG Amsterdam, The Netherlands}

\date{\today}

\begin{abstract}
A steady-state magneto-optical trap (MOT) of fermionic strontium atoms operating on the \RedMotLinewidth-wide~\RedMotTransition~transition is demonstrated. This MOT features \MOTAtomNumberAbstrac, a loading rate of \MOTfluxAbstract, and an average temperature of \MOTAvgtempAbstract. These parameters make it well suited to serve as a source of atoms for continuous-wave superradiant lasers operating on strontium's $\si{\milli \hertz}$-wide clock transition. Such lasers have only been demonstrated using pulsed Sr sources, limiting their range of applications. Our MOT makes an important step toward continuous operation of these devices, paving the way for continuous-wave active optical clocks.
\end{abstract}

\begin{CJK*}{UTF8}{min}
\maketitle
\end{CJK*}

\section{Introduction}
\label{Sec:Introduction}

Optical atomic clocks \cite{LudLow2015ReviewAtomicClock} are proving to be powerful tools to answer fundamental questions in physics. They enable searches for dark matter \cite{DereviankoNature2014, Wcislo2016ExpConstraintDarkMatter, Wcislo2018BoundDarkMatterCoupling}, the study of gravitational waves \cite{Kolkowitz2016GraveWaveClock}, tests of general relativity \cite{Takamoto2020TestGenRelClock, ChouScience2010} or variations in nature’s fundamental constants \cite{Martins2017RevFundamentalConstant, BlattPRL2008, Godun2014ConstraintFundamentalConstant}. In the applied domain they are playing an increasingly important role in geology and geodesy \cite{Bondarescu2015GeophysicalJournalInternational, Mehlstaeubler2018AtomClockGeodesy, Grotti2018NatPhys, McGrew2018Nature}.

Typically, state-of-the-art optical clocks operate in a time-sequential manner, with distinct phases for sample preparation and measurement. This mode of operation introduces aliasing of high-frequency noise, known as the Dick effect \cite{Dick1987OriginalDickEffect}, which limits their performance \cite{Quessada2003DickEffect}. To address this issue, several approaches have been followed: employing high-precision non-destructive detection techniques \cite{Westergaard2010MinimizingDickEffect}, developing novel experimental architectures to achieve high duty cycles \cite{Norcia2019TweezerClock} or interleaving the clock interrogations between two different atomic samples \cite{Takamoto2011BeyondDick, Schioppo2016DualYbClock,Savoie2018InterleavedInterf}. 

A new class of optical clocks based on the principle of superradiant lasing have been proposed as next-generation devices \cite{Chen2005ActiveClock, Meiser2009MilliHzLaser}. Their working principle is inherently suited to continuous operation, thus circumventing the Dick effect. Furthermore, their potential goes beyond addressing this challenge: in the future they could outperform conventional clocks in terms of stability, detection bandwidth, and, by relaxing the requirements on external high-finesse cavities, portability and robustness \cite{Meiser2009MilliHzLaser, Meiser2010SteadyStateSuperradiance}. 
 
Pulsed superradiant lasing has already been demonstrated~\cite{Bohnet2012SuperradiantLaserRb}, in particular using strontium’s \clocktransition~clock transition \cite{Norcia2016SuperradianceStrontiumClock, Norcia2018PRxSuperradianceMHz}. There are ongoing efforts to incorporate steady-state atomic sources with this new type of clock \cite{Bohnet2012SuperradiantLaserRb, iqClock, YbSuperadiance, Muniz2019ThompsonSuperradiantClockSPIE}. While there exist several strontium sources achieving the required flux and temperature \cite{Chen2019Beam, Bennetts2017HighPSDMOT}, all of these operate using bosonic isotopes, due to their simpler atomic structure. However, superradiant lasing using the clock line of bosonic strontium has, so far, not been demonstrated.

For bosonic strontium, the clock linewidth is orders of magnitude smaller than for its fermionic counterpart, and requires high external magnetic fields for the linewidth to be broadened enough for clock interrogation \cite{TaichenachevPRL2006MagneticFieldInduce, OrigliaPRA2018}. Moreover, such fields must feature an exceptional degree of homogeneity and stability across the atomic sample for all atoms to participate in superradiant lasing \cite{iqClockProgressreport}. Conversely, the laser cooling stages that are necessary to achieve the desired temperature and density operate best at low magnetic fields and gradients. These two adverse requirements complicate the integration of a steady-state source of bosonic strontium into an active optical clock. By contrast, a fermionic source seems better suited. Indeed, due to the non-zero nuclear magnetic moment of fermionic strontium, the clock transition naturally opens \cite{LudLow2015ReviewAtomicClock, BoydPHDThesis}, making it accessible without the need for high external magnetic fields.

Here we demonstrate a continuously loaded magneto-optical trap (MOT) of ultracold fermionic \Sr{87} atoms, operated on the \RedMotLinewidth-wide \RedMotTransition~inter-combination line. This narrow-line MOT features \MOTAtomNumber~at an average temperature of \MOTAvgtemp, is loaded with a rate of \MOTflux~and reaches a peak density of \MOTDensity. This steady-state source could be used to provide the high flux of ultracold atoms necessary to refill the gain medium of a steady-state superradiant active clock \cite{Meiser2010SteadyStateSuperradiance, Bohnet2012SuperradiantLaserRb, Norcia2016SuperradianceStrontiumClock}. 

This article is structured as follows. Section~\ref{Sec:Apparatus_overview} briefly describes the experimental setup necessary to produce a continuous MOT of strontium using the \RedMotTransition~transition, while Sec.~\ref{Sec:2D_SWAP_molasses}, \ref{Sec:Slower}, and \ref{Sec:MOT} focus on the specificities pertaining to the fermionic isotope \Sr{87}. Section~\ref{Sec:2D_SWAP_molasses} describes a 2D molasses, making use of the sawtooth-wave adiabatic passage (SWAP) technique to collimate an atomic beam passing from one chamber to the next. Section~\ref{Sec:Slower} describes a narrow-line Zeeman slower adapted to fermionic strontium, thanks to optical pumping into suitable states of the hyperfine manifold of ground-state strontium. Section~\ref{Sec:MOT} characterizes the MOT performance and directly compares the \Sr{88} and \Sr{87} cases. Lastly, Sec.~\ref{Sec:Discussion} discusses the application of this atomic source to superradiant active optical clocks and concludes.

\section{Apparatus overview}
\label{Sec:Apparatus_overview}

In this section we first outline the main components of our apparatus, by describing every laser cooling stage necessary to produce a continuous narrow-line MOT. We then highlight the main differences between steady-state narrow-line MOTs of \Sr{88} (bosonic strontium) and \Sr{87} (fermionic strontium). 

\begin{figure}[tb]
\includegraphics[width=0.98\columnwidth]{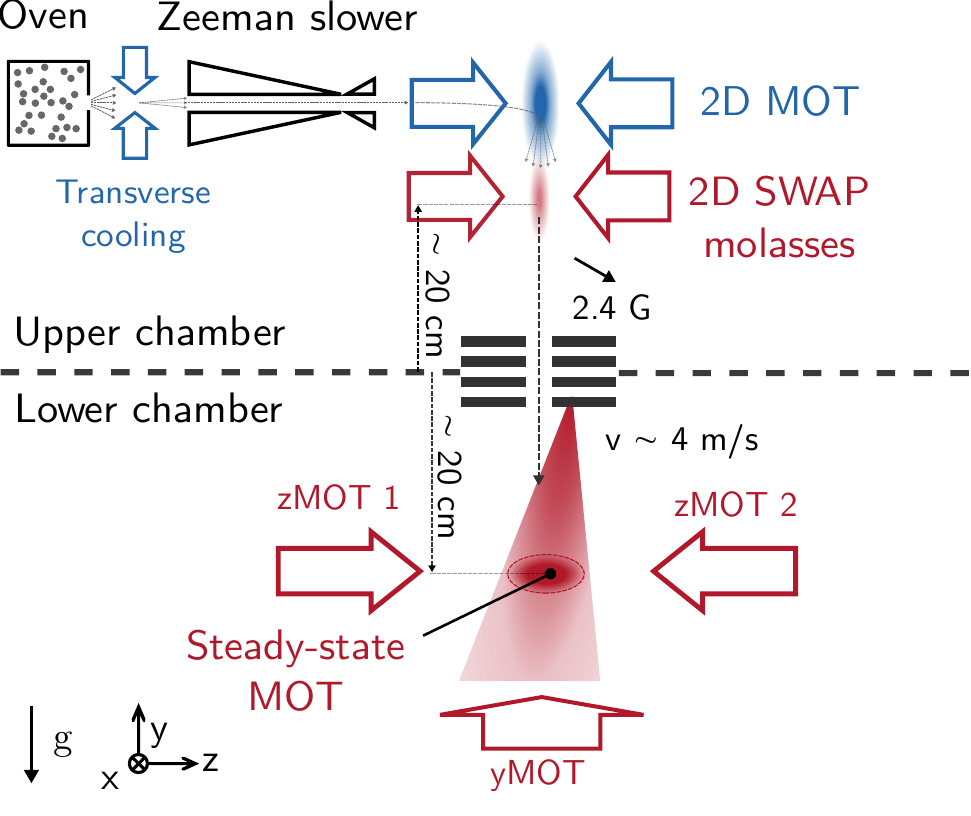}
\caption{\label{fig:Setup_schematic} Schematic of the apparatus producing a steady-state narrow-line MOT of fermionic \Sr{87}. The top chamber features a 2D MOT using the \BlueMotLinewidth-wide \BlueMotTransition~transition, followed by 2D molasses using the \RedMotLinewidth-wide \RedMotTransition~transition. To account for both Doppler broadening and Zeeman shifts in the \Sr{87} ${^3\mathrm{P}_1}$ manifold, we use a variation of the SWAP molasses technique, see Sec.~\ref{Sec:2D_SWAP_molasses}. In the lower chamber where the steady-state narrow-line MOT is produced, we use a dual-frequency scheme to both decelerate and capture the incoming atoms, see Sec.~\ref{Sec:Slower}.}
\end{figure}

Similarly to our previous work on bosonic strontium \cite{Bennetts2017HighPSDMOT}, we stream a beam of atoms through a sequence of spatially separated laser cooling stages, until reaching the $\si{\micro\kelvin}$ regime. Initially, atoms coming from an oven at $\sim \SI{550}{\celsius}$ are slowed down by a Zeeman slower operated on the $\SI{30}{\mega\hertz}$-wide \BlueMotTransition ~transition. For this work, we have upgraded the power of the Zeeman slower beam to improve our atomic flux by more than an order of magnitude, using $\SI{500}{\milli \watt}$ diodes now available from Nichia (Part NDB4916) \cite{Schkolnik2020Nichia500mWDiode}. The high cycling rate of the \BlueMotTransition~transition is necessary to decelerate and cool a significant part of the velocity distribution of the beam exiting the oven. However, parasitic light resonant with this transition, coming, for example, from uncontrolled reflections or fluorescence from the laser-cooled atoms, will be extremely damaging to superradiantly lasing ensembles and quantum degenerate gases \cite{Stellmer2013LaserCoolingToBEC, Chen2020SSBEC, Norcia2016SuperradianceStrontiumClock}. It will also be harmful to the lifetime and temperature of our narrow-line MOT. To address these issues, we opted for a two-chamber design, consisting of an upper and a lower chamber, see Fig.~\ref{fig:Setup_schematic}. We use light addressing \BlueMotTransition~in the upper chamber but not in the lower one. 
Indeed, only light on the much narrower, $\SI{7.48(5)}{\kilo \hertz}$ wide \cite{Nicholson2015Clock10PowMinus18, Lu2017Linewidth88Sr3P0}, \RedMotTransition ~transition has been shown to be compatible with Bose-Einstein condensates \cite{Stellmer2013LaserCoolingToBEC, Chen2020SSBEC}.

In the upper chamber, Zeeman slowed atoms are gathered in a 2D MOT using the \BlueMotTransition ~transition, see Fig.~\ref{fig:Setup_schematic}. This 2D MOT produces a vertical, downward propagating atomic beam with a radial velocity $\lessapprox \SI{0.5}{\meter \per \second}$.
In order to efficiently transfer atoms to the lower chamber through an $\SI{8}{\milli \meter}$-diameter baffle, this atomic beam requires collimation. In typical double-chamber experiments, this collimation can be achieved by accelerating the atoms exiting the 2D MOT to high velocities. However due to the low scattering rate of the \RedMotTransition~transition, we would not be able to stay below the capture velocity in the lower chamber. Thus, in order to achieve a high-transfer efficiency we need to radially cool the beam. We achieve this goal by employing a 2D molasses operating on the \RedMotTransition~transition and located directly below the 2D MOT.

Due to gravity, atoms accelerate and reach the center of the lower chamber, located $\sim \SI{40}{\centi \meter}$ below the 2D MOT, with a downward velocity distribution peaked at $\sim \SI{4}{\meter\per\second}$. A counter-propagating laser beam slows the atoms over a $\sim \SI{5}{\centi \meter}$ distance using the \RedMotTransition ~transition. Finally, a narrow-line 5-beam 3D MOT using the same transition captures the atoms. The vertical, upward propagating beam serves a dual purpose, both slowing the atoms coming from the upper chamber and acting as a MOT beam. Once in the MOT, atoms are slow enough that gravity is sufficient to provide the downward restoring force.

The above description applies to our previous work on \Sr{88} \cite{Bennetts2017HighPSDMOT} and to our new work on \Sr{87} described here. However, in order to deal with the internal structure of \Sr{87}, our new architecture implements two key changes. Firstly, the collimating 2D molasses needs to be upgraded using the SWAP technique \cite{Norcia2018TheoSWAP}, to account for the magnetic sensitivity of all \Sr{87} $^3\mathrm{P}_1$ Zeeman sub-levels, see Sec.~\ref{Sec:2D_SWAP_molasses}. Secondly, optical pumping into suitable states of the Sr ${^1\mathrm{S}_0}$ hyperfine manifold is necessary to ensure efficient slowing of falling atoms in the lower chamber. We use additional laser light addressing the \PumpingTransition~transition \cite{Katori1999SrMOTRecoilTemp} to perform optical pumping thus enhancing the atomic flux into the MOT, see Sec.~\ref{Sec:Slower} and \ref{Sec:MOT}.

\section{2D SWAP molasses}
\label{Sec:2D_SWAP_molasses}

In this section, we describe how the 2D molasses using the narrow \RedMotTransition~transition collimates the atomic beam in order for the atoms to pass from the upper chamber to the lower chamber. We first detail how we previously achieved this collimation for bosonic \Sr{88}, then highlight the additional challenges pertaining to fermionic \Sr{87}. We resolve these challenges using SWAP cooling, a technique which uses sawtooth-shaped detuning ramps of the laser cooling light. We recall two criteria for this technique to work. We then explore its limits in our implementation and show that our 2D molasses performs better using SWAP laser detuning ramps rather than inverted SWAP or symmetric ramps. Finally, we measure and model the cooling efficiency of the 2D molasses in dependence of the laser powers and ramp durations. 

\subsection{SWAP molasses needed for fermionic Sr}
\label{SubSec:SWAP_for_fermions}
The collimation molasses is located just below the broad-line 2D MOT that uses the \BlueMotTransition~transition. It exploits the much lower Doppler temperature limit of the narrow-line \RedMotTransition~transition to further reduce the radial velocity spread of the downward propagating atomic beam, see Fig.~\ref{fig:Setup_schematic}. As atoms fall with a velocity of $\sim \SI{3}{\meter \per \second}$, radial cooling with this 2D molasses reaches steady state in at most $\SI{10}{\milli \second}$. To achieve a sufficiently long interaction time the molasses beams are elliptical, with a $1/e^2$ diameter of $\SI{45.6}{\milli \meter}$ along the vertical axis and $ \SI{18.2}{\milli \meter}$ along the horizontal axis.

The molasses needs to address the atoms efficiently, despite a Doppler broadening about 100 times larger than the transition's natural width $\gamma = 2 \pi \times \SI{7.5}{\kilo \hertz}$. For bosonic strontium we therefore broaden the molasses beams' frequency spectrum, forming a frequency comb spanning from $30$ to $\SI{750}{\kilo \hertz}$ to the red of the transition with a $\SI{25}{\kilo \hertz}$ spacing \cite{Bennetts2017HighPSDMOT}. An additional challenge arises from the presence of a residual 2D quadrupole magnetic field with a gradient of $\SI{2}{\gauss \per \centi \meter}$ at the location of the molasses. This field originates from the 2D quadrupole field of the nearby 2D MOT and it cannot be completely removed. This residual field leads to Zeeman broadening of the magnetic \RedMotTransition~transitions and position dependent light scattering forces. This challenge is overcome by addressing only the $|\mathrm{J} = 0, m_{\mathrm{J}} = 0 \rangle - |\mathrm{J}' = 1, m_{\mathrm{J'}} = 0 \rangle$, $\pi$, non-magnetic transition of \Sr{88}. To this end, we add a vertical bias field of $\SI{1.2}{\gauss}$ to Zeeman shift the other two, $\sigma^{\pm}$, transitions away from resonance. At the location of the atomic beam, the average field amplitude is $\SI{2.4}{\gauss}$, creating a Zeeman shift of about $\SI{5}{\mega \hertz}$, a significant part of the field being along the $z$ axis due to displacement between the atomic beam and the zero of the 2D quadrupole field.

\begin{figure}[tb]
\includegraphics[width=0.98\columnwidth]{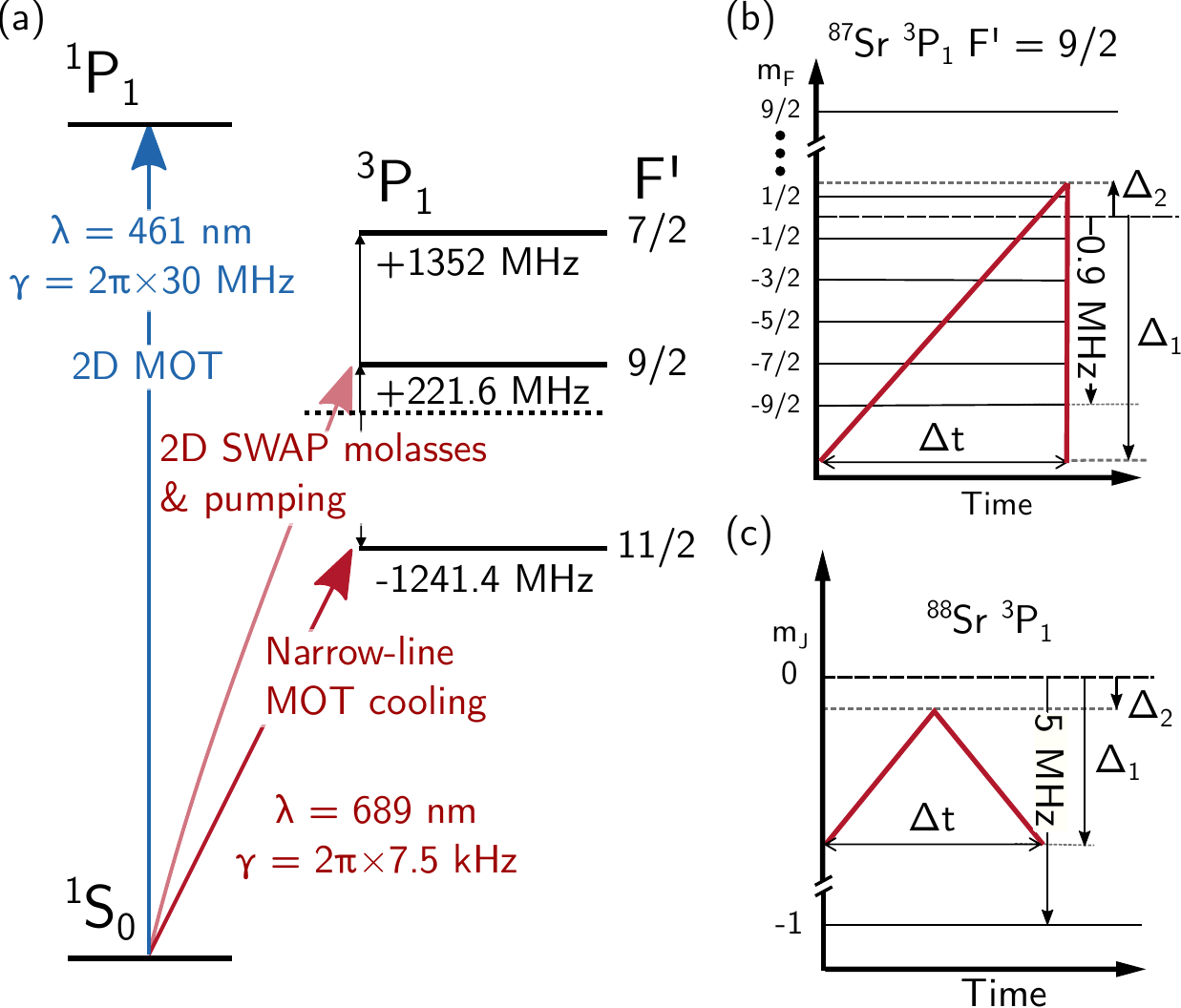}
\caption{\label{fig:87Srstructure} (a) \Sr{87} energy levels relevant for our experiment. The arrows indicate all transitions necessary to obtain our steady-state narrow-line MOT. The hyperfine shifts of the ${^3\mathrm{P}_1}$ manifold are referenced to the single ${^3\mathrm{P}_1}$ state of \Sr{88}. (b) Magnetic sub-levels of the $^3\mathrm{P}_1 \, \mathrm{F'} = 9/2$ state for an atom at rest in presence of a field of $\SI{2.4}{\gauss}$. Also shown is a schematic of the detuning sweep (red lines) used by the 2D SWAP molasses to cool and collimate the atoms propagating toward the lower chamber. (c) Schematic of the symmetric sweep used to collimate bosonic atoms \cite{Bennetts2017HighPSDMOT}. The ramp parameters shown are the sweep duration $\Delta t$, the laser start detuning $\Delta_{1}$ and stop detuning $\Delta_{2}$.}
\end{figure}

\begin{figure}[tb]
\includegraphics[width=0.98\columnwidth]{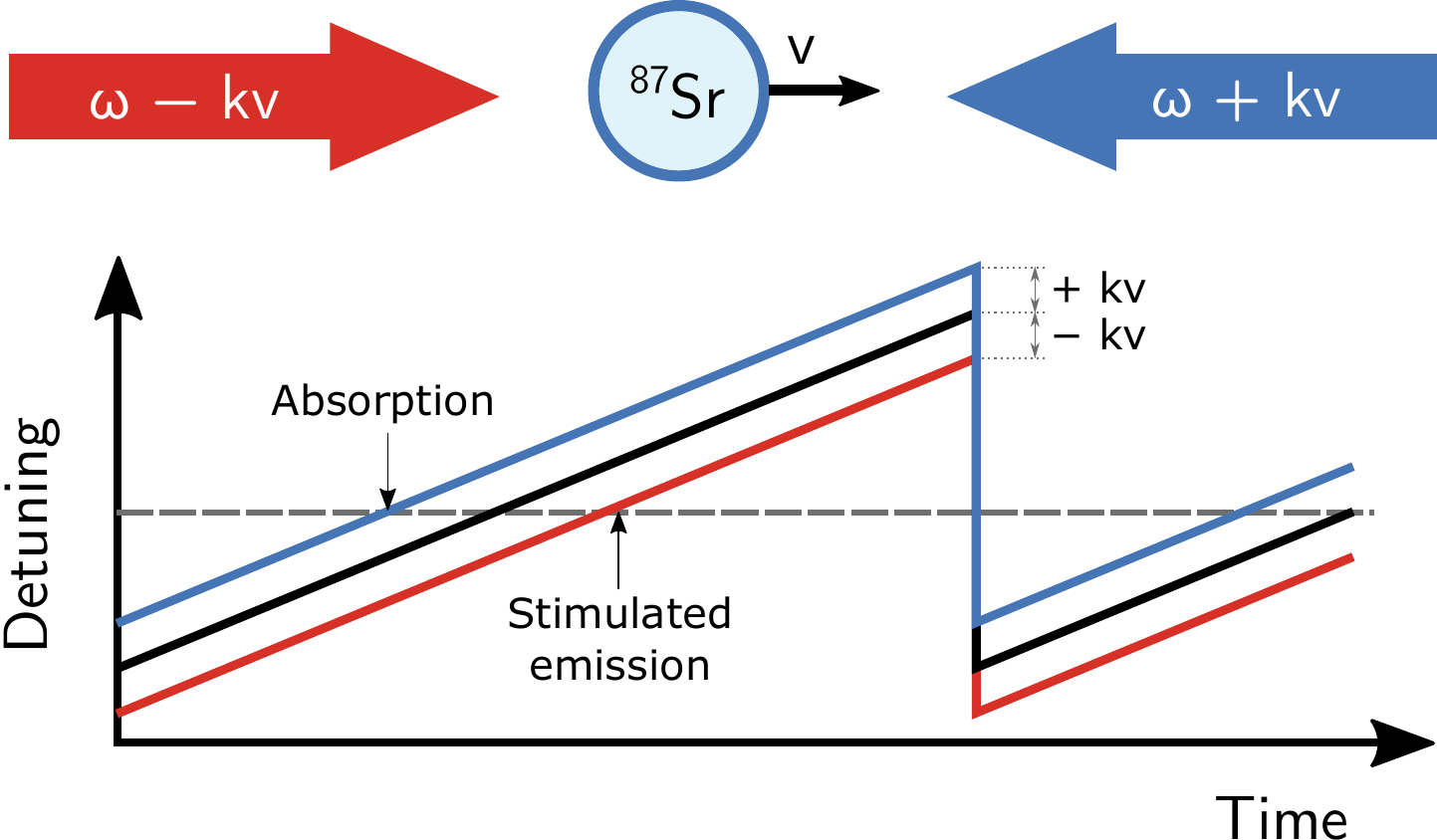}
\caption{\label{fig:SWAP_principle} SWAP working principle. The plain black trace describes the laser frequency sweep in the laboratory frame. In the reference frame of the atom, the detuning of each of the two counter-propagating beams (blue and red traces) is Doppler shifted by $\vec{k}\cdot\vec{v}$. This shift introduces an effective time ordering, ensuring that the momenta transferred to the atom are oriented against its direction of travel.}
\end{figure}

Due to the hyperfine structure of \Sr{87}, non-magnetic transitions from $^1\mathrm{S}_0$ to $^3\mathrm{P}_1$ do not exist, see Fig.~\ref{fig:87Srstructure}(a). Therefore, the method just described no longer provides efficient transfer of atoms to the lower chamber. To overcome this difficulty we use SWAP cooling \cite{Norcia2018TheoSWAP, Bartolotta2018SWAP, Muniz2018RobustSWAPMOT, Snigirev2019SWAPSrMOT, Petersen2019DySWAP, Greve2019SWAPRaman}, a technique that uses two rapid adiabatic passages to transfer atoms to a long-lived excited state and back to the ground state. In this work these states are $^3\mathrm{P}_1$ and $^1\mathrm{S}_0$ respectively. SWAP cooling exploits the time ordering of the adiabatic passages, transferring photons between two counter-propagating laser beams, and ensuring that the recoil momenta associated with photon absorption and stimulated emission are both oriented against the atoms' direction of travel, see Fig.~\ref{fig:SWAP_principle}. SWAP cooling has the advantage of robustly addressing a wide range of effective detunings, while increasing deceleration. Therefore it can be used on atomic samples whose Doppler broadening is much wider than the \RedMotTransition~transition's width, and it can address several transitions with distinct differential Zeeman shifts in a single sweep. 

\subsection{SWAP cooling criteria}
\label{SubSec:SWAP_criterion}

For SWAP cooling to be effective two main criteria must be fulfilled \cite{Norcia2018TheoSWAP, Bartolotta2018SWAP}. The first is the adiabaticity of the rapid adiabatic passage from one internal state to the other, which is required for both absorption and stimulated emission. To be adiabatic, the square of the Rabi frequency of the laser light $\Omega^{2}$ needs to be high compared to the laser detuning sweep rate $\alpha $, i.e. $\kappa = \Omega^{2} / \alpha \gg 1 $. Here $\alpha =\Delta \omega / \Delta t$, $\Delta \omega$ is the full detuning range spanned, and $\Delta t$ is the duration of a single sweep. Indeed, for a Landau-Zener type crossing the probability of transfer is given by $P_{ \mathrm{LZ}} = 1 - \exp \left( -\pi\kappa /2 \right)$ \cite{Zener1932LandauZenerCrossing, Bartolotta2018SWAP}. 

The second criterion concerns the time an atom must spend in the excited state in order to perform a complete cycle of absorption and stimulated emission. This time must be smaller than the natural decay time by spontaneous emission $1/\gamma$. If this criterion is not met, spontaneous emission will occur before the stimulated emission, reducing the cooling efficiency. Moreover, if the atom absorbs a second photon by adiabatic passage from the co-propagating beam, momentum will be imparted in the wrong direction. This second criterion is expressed by $\tau_{e} = | 2 (k v - 2 \omega_r) / \alpha | \ll 1/\gamma$. Here $v$ is the velocity of the atom, and $k$ and $\omega_r = \hbar k^2 / 2 m$ are the wavevector and the recoil frequency associated with the \RedMotTransition~transition, where $\hbar$ is the reduced Planck constant and $m$ is the atom's mass. For high-velocity atoms, where $k v \gg 2 \omega_r$, the second criterion requires the sweep rate $\alpha$ to be fast, while maintaining the first, adiabaticity criterion $\kappa \gg 1$. 
 
Due to the presence of a magnetic field, the detuning range $\Delta \omega$ must be large enough to address all \Sr{87} Zeeman sub-levels. However, the available laser power does not allow us to fulfill the adiabaticity criterion $\kappa$ for a large range $\Delta \omega$. To mitigate this issue we use the \PumpingTransition ~transition, because it has the lowest differential magnetic moment of all the possible hyperfine transitions. Additionally, we further reduce the necessary $\Delta \omega$ by offsetting the center detuning of the sweep to the red of the unperturbed transition, see Fig.~\ref{fig:87Srstructure}(b). In this way we compromise between executing a well performing SWAP molasses for the most populated \mf~states and addressing as many states as possible.

\subsection{SWAP and inverted SWAP working regimes}
\label{SubSec:SWAP_inverted_SWAP}

\begin{figure}[tb]
\includegraphics[width=0.98\columnwidth]{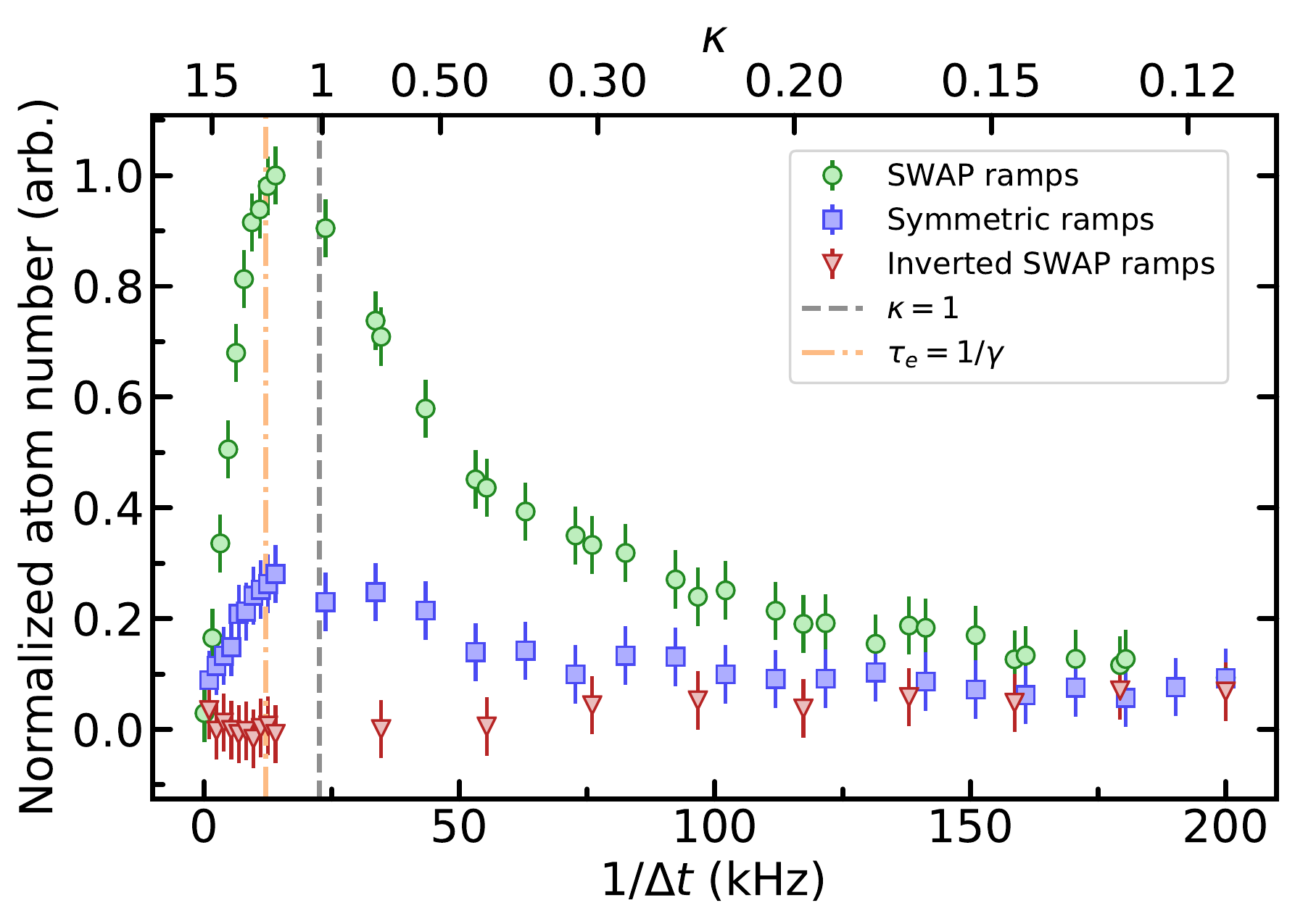}
\caption{\label{fig:Swap_symetry} Relative number of atoms captured in the narrow-line 3D MOT for three different ramp types with varying sweep frequency $1/\Delta t$ and fixed MOT loading time of $\SI{8.5}{\second}$. The green disks, blue squares, and red triangles show respectively SWAP, symmetric, and inverted SWAP ramps. The uncertainty on the data is estimated by performing 16 measurements at a value of $1/\Delta t = \SI{14}{\kilo \hertz}$ using SWAP ramps, and we give the resulting standard deviation as error bars for all points. The gray dashed line and the orange dashed-dotted line show the conditions $\kappa =1 $ and $\tau_{e} = 1/\gamma$ ($v = \SI{0.2}{\meter\per\second}$), respectively. We use the maximum available power for the 2D red molasses beams, which is $\SI{14.2}{\milli \watt}$ for one beam pair and $\SI{15.2}{\milli \watt}$ for the other. This corresponds to a Rabi frequency averaged across all transitions of about \OmegaSWAPAverage. The molasses light detuning is modulated as shown in Fig.~\ref{fig:87Srstructure}, between $\Delta_1 = - 2 \pi \times \SI{2}{\mega \hertz} $ and $\Delta_2 = + 2 \pi \times \SI{0.2}{\mega \hertz}$ for SWAP and symmetric ramps and between $\Delta_1 = - 2 \pi \times \SI{0.2}{\mega \hertz}$ and $\Delta_2 = - 2 \pi \times \SI{2}{\mega \hertz}$ for inverted SWAP ramps. Note that the MOT atom number is not strictly proportional to the flux arriving in the lower chamber, due to possible optical pumping processes and other saturation effects (see~Sec.~\ref{Sec:Slower}).
}
\end{figure}

For SWAP cooling, the sign of the sawtooth ramp slope determines whether the time ordering associated with absorption and stimulated emission results in net momentum gain or loss. A positive slope (SWAP ramps), such as the one shown in Fig.~\ref{fig:SWAP_principle}, results in deceleration and cooling, while a negative slope (inverted SWAP ramps) results in acceleration and heating. \cite{Petersen2019DySWAP,Bartolotta2018SWAP}. In Fig.~\ref{fig:Swap_symetry} we compare the atom number loaded into the MOT (see Sec.~\ref{Sec:MOT}) for molasses following SWAP, inverted SWAP, and symmetric ramps, with varying sweep frequency $1/\Delta t$. For all explored $\Delta t$, the flux of loaded atoms is the highest using SWAP ramps, while inverted SWAP ramps perform the worst. This difference is the most pronounced at $1/ \Delta t_{\mathrm{opt}} \sim \SI{15}{\kilo\hertz}$, corresponding to an adiabaticity criterion of $\kappa \sim 1.5$, while in the limits of short or long sweep durations the fluxes become similar. This comparison of the three types of ramp shows that the time ordering of the molasses beams' action matters and that the adiabaticity of the emission and absorption of the rapid adiabatic passages plays a significant role here.

The results of Fig.~\ref{fig:Swap_symetry} can be explained by considering the two SWAP criteria previously given. Indeed, for short durations the adiabaticity parameter decreases to $\kappa \ll 1$ and the probability of stimulated emission or absorption becomes negligible. This reduces the cooling efficiency and effectively breaks the time ordering. In this limit the cooling efficiencies of SWAP and inverted SWAP ramps converge and result in similar fluxes. For $\kappa \gg 1$, corresponding to long durations, the adiabaticity criterion is fulfilled, but the second criterion is broken, as $\tau_{e} \propto 1 / | \alpha | \propto \Delta t $. For $\tau_{e} \gg 1/\gamma$ the cooling reduces to two independent stimulated absorption events each followed by spontaneous emission, one from the counter-propagating beam and one from the co-propagating beam. In this situation the average slowing force approaches zero for all three ramp types.

\subsection{Modeling SWAP molasses}
\label{SubSec:Modeling_SWAP}

\begin{figure}[tb]
\includegraphics[width=0.98\columnwidth]{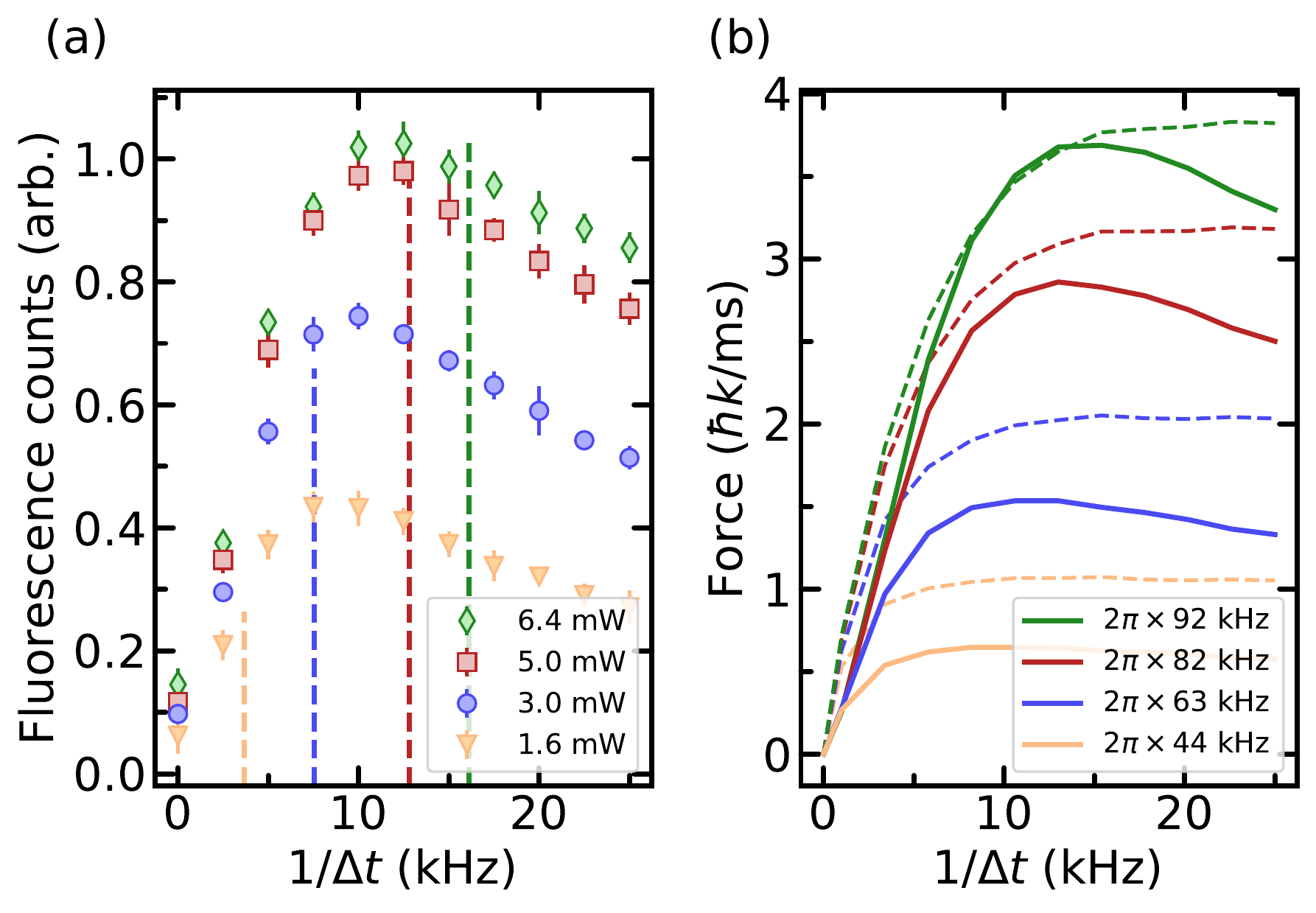}
\caption{\label{fig:Swap_omega_vs_frequency} Molasses performance characterization. (a) Experimentally measured transfer efficiency into the bottom chamber for several beam powers with varying sweep frequency $1/\Delta t$. The powers stated in the legend refer to the average laser power for each beam. Each data point represents the average of ten independent measurements and the error bars show the standard deviation of these measurements. The vertical dashed lines highlight the location of $\kappa = \Omega^{2}/ \alpha = 1.5 $. (b) Numerical estimations of the molasses' average restoring force exerted on an atom exiting the 2D MOT, with the same parameters as in (a). The curves represent thermal averages over a 1D Maxwell Boltzmann distribution at $\SI{1}{\milli \kelvin}$ for two regimes, one integrated only over the low-velocity class $|v| < \SI{0.25}{\meter \per \second}$ (solid line) and the other integrated over the high-velocity class $|v| > \SI{0.25}{\meter \per \second}$ (dashed lines). The given Rabi frequencies result from spatial averages of the intensity across a $1/e^{2}$ beam waist.}
\end{figure}

To explore the power dependence of the 2D SWAP molasses, we measure the relative transfer efficiency to the lower chamber for different laser beam powers and sweep durations. These results are shown in Fig.~\ref{fig:Swap_omega_vs_frequency}(a).

We obtain this data by collecting fluorescence from the falling atoms in the lower chamber. For this purpose, we illuminate the passing atoms with light addressing the fast cycling \BlueMotTransition~transition, with no other beam present in the lower chamber. This detection method provides a more direct measure of the atomic flux arriving into the lower chamber that what is given by the MOT atom number.

For each chosen power we observe the existence of an optimal sweep frequency close to $1 / \Delta t_{\mathrm{opt}} \sim \SI{10}{\kilo\hertz}$, that shows only a small dependence with the laser beam power. The location of this optimum does not quite correspond to the value of the adiabaticity criterion $\kappa$, indicated in Fig.~\ref{fig:Swap_omega_vs_frequency}(a) by vertical dashed lines. This mismatch might arise from the laser beams' finite interaction time with the atoms. Indeed, in order to pass through the $\SI{8}{\milli \meter}$-diameter baffle between the two vacuum chambers, the radial velocity must be reduced below $v < v_{\mathrm{min}} \approx \SI{0.07}{\meter \per \second}$. However, atoms exiting the 2D MOT have a radial velocity up to $\SI{0.5}{\meter \per \second}$. Successful passage to the lower chamber therefore requires that the 2D molasses removes about 50 recoil momenta per beam pair. Assuming a momentum transfer of one photon recoil per sweep, achieving this deceleration within the finite interaction time with the laser beams would require a minimum sweep frequency of $1 / \Delta t \sim \SI{3}{\kilo \hertz}$. The measured optimum sweep frequency is three times slower, indicating that on average only $0.3$ photons might be transferred per sweep, corresponding to a restoring force of $\sim \SI{3}{\hbar \textit{k} \per \milli \second}$.

To deepen our insight into SWAP molasses cooling, we theoretically estimate the average restoring force exerted on the atoms exiting the 2D MOT, see Fig.~\ref{fig:Swap_omega_vs_frequency}(b). We limit our discussion to the average force exerted by the molasses, as reproducing the data shown in Fig~\ref{fig:Swap_omega_vs_frequency}(a) would require introducing free parameters into our model that we do not have access to. Such parameters include the atomic distribution exiting the 2D MOT, the transit time and the maximum transfer velocity of the atoms. Moreover, including them in the model would not provide deeper insight into the molasses cooling process itself. 

The results of Fig.~\ref{fig:Swap_omega_vs_frequency}(b) are obtained by expanding on the theory developed by \citet{Bartolotta2018SWAP} to take into account the internal structure of \Sr{87}. In particular, we numerically integrate the optical Bloch equations for the sweep durations and Rabi frequencies shown in Fig.~\ref{fig:Swap_omega_vs_frequency}(b) using the finite set of states,
\begin{equation*}
 \ket{m_{\mathrm{F}}, p + n \hbar k}, \ket{m_{\mathrm{F^{\prime}}}, p + n \hbar k},
\end{equation*}
where $n \in \{ 2 , 1 , 0 , -1 , -2 \}$, $p$ defines the starting momentum used in the calculation, and we have used short-hand notation for a particle with momentum $p + n \hbar k$ in the \groundState~$m_{\mathrm{F}}$ and \StirringState~$m_{\mathrm{F^{\prime}}}$ states respectively.

We solve these equations stepping through several starting momenta, $p/\hbar k \in \{ 70 , 69.5 , 69 ,\ldots , 0\} $ and one sweep cycle of duration $\Delta t$ to obtain a momentum dependent force $ F(p) = (\mathrm{tr}(\rho_{i,p} \hat{p}) - \mathrm{tr}(\rho_{f,p} \hat{p})) /\Delta t $. Here $\hat{p}$ is the atomic momentum operator, $\rho_{i,p} = \sum\limits_{m_{\mathrm{F}}}{} P_{i, m_{\mathrm{F}} , p} \ket{m_{\mathrm{F}}, p}\bra{m_{\mathrm{F}}, p}$ is the initial state used in the calculation and $\rho_{f,p}$ is the final state after the single sweep.

To take into account optical pumping, the initial distribution over internal states $m_{\mathrm{F}}$ described by $P_{i, m_{\mathrm{F}}, p}$ in $\rho_{i}$ is obtained by an iterative process. Starting from a fully depolarized state with momentum $70 \hbar k$, i.e. $P_{i, m_{\mathrm{F}} , 70 \hbar k} = 1/10$, the final occupation of the internal states of each momentum step is used as an input occupation for the following step, i.e. $P_{i, m_{\mathrm{F}} , p^{\prime} } = \sum\limits_{n}{} \bra{m_{\mathrm{F}},p + n\hbar k} \rho_{f,p} \ket{m_{\mathrm{F}},p+ n\hbar k}$, with $p - p^{\prime} = 0.5 \hbar k$ corresponding to our choice of discretization. We checked that the result of the calculation depends weakly on this somewhat arbitrary choice and that the shape of the curves shown in Fig.~\ref{fig:Swap_omega_vs_frequency} (b) remains similar when using a grid with $p - p^{\prime} = \{0.5,1,2\} \hbar k$.

The curves shown in Fig.~\ref{fig:Swap_omega_vs_frequency}(b) average the force obtained over a 1D Maxwell-Boltzmann distribution split into two velocity classes, $|v| < v_{c}$ and $|v| > v_{c}$. Atoms in the low-velocity class are rapidly deexcited via stimulated emission, whereas atoms in the high-velocity class remain unaddressed long enough to spontaneously emit a photon. Here we choose to distinguish these two classes for $1 /\Delta t=\SI{15}{\kilo \hertz}$, which gives $v_{c} = \SI{0.25}{\meter \per \second}$ using the second SWAP cooling criterion, see Sec.~\ref{SubSec:SWAP_criterion}.

Our theoretical results confirm that for the high-velocity atoms, $v>v_{c}$, the cooling mostly reduces to a single absorption by adiabatic passage from the counter-propagating beam, followed by spontaneous emission. Low velocity atoms $v <v_{c}$ have an effective detuning dominated by the Zeeman shifts. Since our sweep range is red detuned from the unperturbed transition, it leaves the magnetic sub-levels with $m_{\mathrm{F}} > \frac{3}{2}$ unaddressed. However, our simulations show that optical pumping is optimal for a sweep duration corresponding to $\sim \SI{10}{\kilo \hertz}$. This optical pumping enhances the population of the states that are efficiently addressed by the molasses beams, particularly those with low $|m_{\mathrm{F}}|$ values. Failing to take the effect of optical pumping into account shifts the optima shown in Fig.~\ref{fig:Swap_omega_vs_frequency}(b) to significantly lower sweep duration.

From these simulations we obtain the maximum average restoring force $F_{\mathrm{max}} = \SI{3.8}{\hbar \textit{k} \per \milli \second}$ provided by the SWAP molasses. This relatively small force is enough to collimate most atoms such that they pass through the $\SI{8}{\milli \meter}$-diameter hole between chambers, and is consistent with the rough estimation $\sim \SI{3}{\hbar \textit{k} \per \milli \second}$ we gave previously.

To finalize this section, we summarize the SWAP molasses cooling parameters used in the remainder of this work. The optical powers are \PowerLowPowerBeam~for one pair of beams and \PowerHighPowerBeam~for the other. Their polarization is linear and perpendicular to the $y$ axis. The sweep duration is $\Delta t =$~\SweepTime, while the molasses detuning is modulated between $\Delta_1 = - 2 \pi \times \SI{2}{\mega \hertz} $ and $\Delta_2 = + 2 \pi \times \SI{0.2}{\mega \hertz}$. Based on our experimental explorations, we find this set of parameters to provide optimal flux to our narrow-line 3D MOT.

\section{Laser slowing stage}
\label{Sec:Slower}

Atoms flowing through the 2D SWAP molasses region can be captured in the lower chamber by a narrow-line MOT operating on the \RedMotTransition~transition, see Fig.~\ref{fig:Setup_schematic}. Because of the limited deceleration achievable with the \RedMotTransition~transition, we slow the atoms before they reach the MOT. In the following, we first describe the operation of this slowing stage for bosonic isotopes. Then we highlight the additional challenges for the fermionic isotope. We finish this section by describing the frequency ramps used to efficiently slow the atoms.

Accelerated by gravity, atoms reach the entrance of the lower chamber with a downward velocity distribution peaked at $\sim \SI{4}{\meter \per \second}$, which exceeds the MOT's capture velocity. Therefore we employ a counter-propagating laser beam addressing the \RedMotTransition~transition to begin slowing the atoms before they reach the MOT, which is located $\sim \SI{40}{\centi \meter}$ below the 2D SWAP molasses. This beam is circularly polarized to address the $\sigma^{+}$ transition and its frequency is red detuned from the unshifted resonance.

\begin{figure}[tb]
\includegraphics[width=0.98\columnwidth]{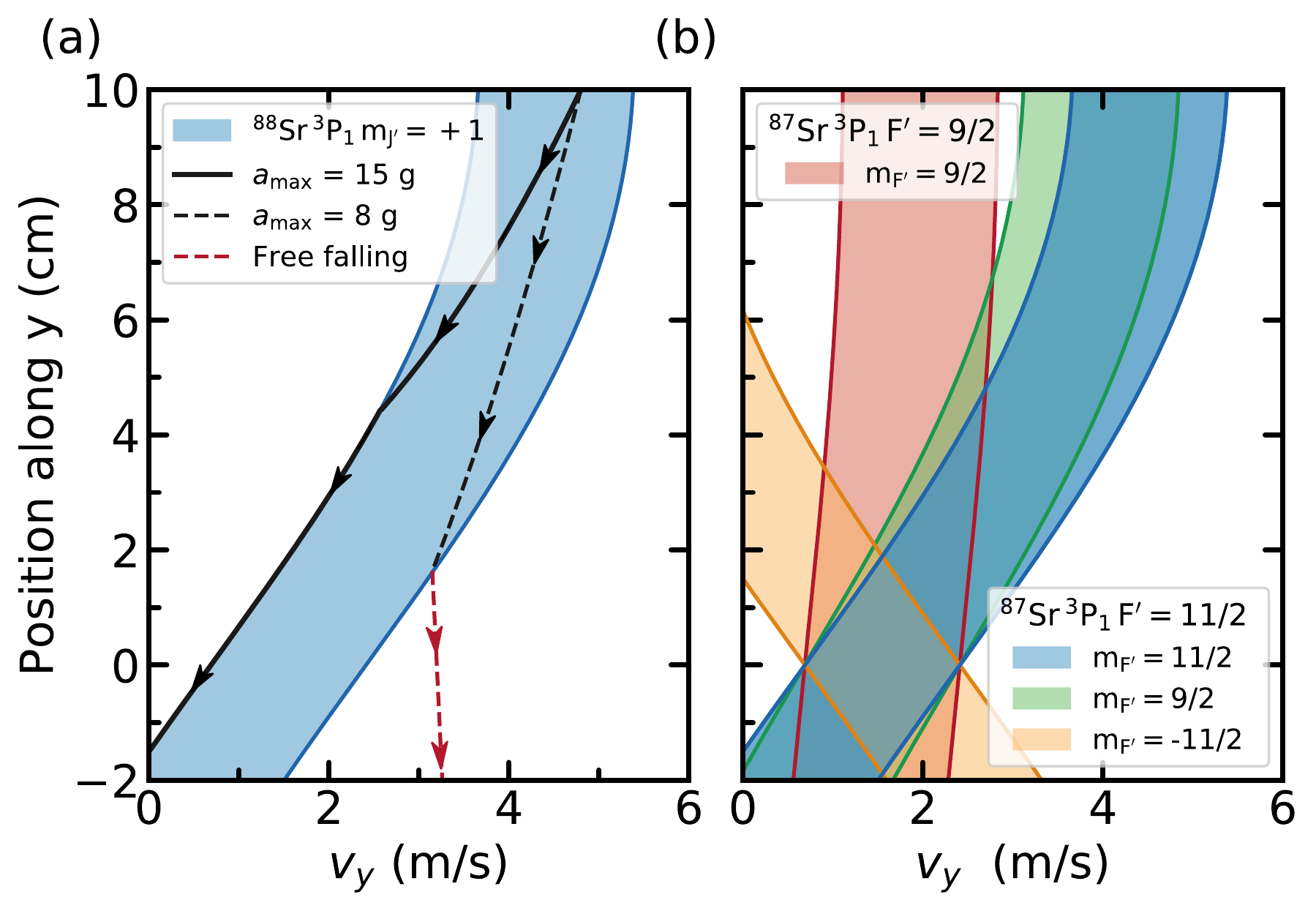}
\caption{\label{fig:Slower_Stage_Phase_Space_Plot} Phase-space plots of falling atoms under the influence of the slower beam. The highlighted regions show the resonance windows of the various \RedMotTransition~transitions, defined by $\Delta_{2} < \delta_{Z}(B(y)) - \delta_{D}(v_{y}) < \Delta_{1} $. Here $\delta_{D}(v_{y})$ is the Doppler shift, $\delta_{Z}(B(y))$ the position-dependent Zeeman shift resulting from the quadrupole field of the MOT, and $\Delta_{1}$ and $\Delta_{2}$ are the slower laser start and stop frequency detunings given in Table~\ref{table:CaptureMOT_parameters} (YMOT). The lines show falling trajectories for two different decelerations. Atoms not efficiently slowed (dashed lines) can fall out of the resonance window and be lost. (a) For bosonic isotopes, such as \Sr{88}, as the falling atoms slow down (black solid line), the change of Doppler shift is partially compensated thanks to the magnetic field gradient from the quadrupole, whose center is at $y=0$. (b) For fermionic \Sr{87}, due to it's internal structure, the capture velocity depends on the hyperfine transition being addressed. In particular, the cooling transition has the highest possible capture velocity, while the pumping transition has a much smaller capture velocity due to the smaller Land\'e factor. For clarity, not all of the possible transitions are shown.}
\end{figure}

For bosonic strontium, the slowing stage operates as a Zeeman slower and its working principle is sketched in Fig.~\ref{fig:Slower_Stage_Phase_Space_Plot}(a). As atoms decelerate, the change in Doppler shift $\delta_{\mathrm{Doppler}}$ is partially compensated, thanks to the vertical magnetic field gradient produced by the MOT, $dB/dy = \SI{0.32}{\gauss \per \centi \meter}$. This gradient is mostly linear up to $\SI{10}{\centi \meter}$ away from the MOT quadrupole center, beyond which the magnetic field amplitude decreases. Due to the high initial atom velocity the Zeeman shift alone is insufficient to fully address the atoms, so we also modulate the laser detuning with a span of $\Delta \omega_{\mathrm{ZS}}$ = $2 \pi \times \SI{4.05}{\mega \hertz}$, thus forming a ``white light" slower \cite{Zhu1991FirstExpWhiteLightSlowing}. 

The choice of laser detuning $\Delta_{2}$, together with the maximum magnetic field experienced by the falling atoms $B_{\mathrm{max}}$, defines a capture velocity for the slowing stage, $v_{\mathrm{capture}} = (\delta_{Z}(B_{\mathrm{max}}) - \Delta_{2})/k$, see Fig.~\ref{fig:Slower_Stage_Phase_Space_Plot}.  For bosonic strontium $v_{\mathrm{capture}} = \SI{5.3}{\metre \per \second}$, faster than the velocity at which the atoms fall of $\sim \SI{4}{\meter \per\second}$. For fermionic strontium, the hyperfine structure complicates the slowing stage, see Fig.~\ref{fig:Slower_Stage_Phase_Space_Plot}(b). From all the hyperfine transitions (see Fig.~\ref{fig:87Srstructure}) only the \CoolingTransition~``cooling" transition can have high enough capture velocity to slow the atoms while being compatible with the MOT configuration described in Sec.~\ref{Sec:MOT}. Therefore we use this transition in order to slow atoms entering the lower chamber. Note that the condition $v_{\mathrm{capture}} > \SI{4}{\meter\per\second} $ can only be fulfilled for atoms in the $m_{\mathrm{F}} > \frac{3}{2}$ states, were $\vec{y}$ defines the quantization axis. Without optical pumping, fast atoms entering the lower chamber in other states will be lost. This loss mechanism can be particularly relevant as our simulations show that the collimating 2D molasses in the upper chamber results in a mixture of several $m_{\mathrm{F}}$ states, predominantly with low $|m_{\mathrm{F}}|$ values.

To optically pump ground-state atoms to the $m_{\mathrm{F}} > 3/2$ Zeeman sub-levels, we use dual-frequency laser light for the vertical slower beam, addressing two different hyperfine manifolds. Firstly, we use the cooling transition, because of its high capture velocity. Secondly, we use the \PumpingTransition~``pumping" transition to optically pump atoms that would otherwise remain unaddressed. For atoms entering the lower chamber, the slower light is typically off-resonant by a few $\si{\mega \hertz}$. However, atoms are illuminated by the vertical beam for $\sim \SI{10}{\centi \meter}$ traveling distance before reaching the region where the slowing starts. For velocities of $2-\SI{4}{\meter\per\second}$, the resulting interaction times of $50-\SI{25}{\milli \second}$ allow for off-resonant scattering of about $1-100$ photons, thus optically pumping atoms to addressed states.

As atoms continue to slow down, the $\sigma^{+}$ polarized slower beam keeps the atoms in the $m_{\mathrm{F}} = 9/2$ magnetic sub-level. However, atoms traveling toward the MOT location can experience spin flips relative to the quantization axis $\vec{y}$ if they are able to adiabatically follow the local magnetic field. For these atoms the deceleration provided by the cooling transition will be severely reduced.

Indeed, such a spin flip is possible. Close to the quadrupole center slowed atoms have a velocity $v_y \gtrsim \SI{0.5}{\meter \per \second}$. These falling atoms will experience a time-dependent magnetic field that rapidly changes direction. If the rate of magnetic field change is slower than the Larmor frequency of the \groundState~state then the probability of a spin flip is high. This criterion can be expressed as  $\chi = \frac{1}{B}\frac{dB}{dt}/\gamma_I B \ll 1$, where $\gamma_{I} =$~\GroundStateGyromagneticRatio is the gyromagnetic ratio of the \groundState~state. 

The criterion $\chi = 1$ delimits a horizontal disk with radius $r \approx \sqrt{v_y /(\gamma_I \, dB/dy)} \approx \SI{4}{\milli \meter}$. This radius is comparable to the atomic beam radius, apertured by the $\SI{8}{\milli \meter}$-diameter baffle between chambers. Therefore some atoms will fall outside the disk and their spin will adiabatically follow the local magnetic field. Relative to the quantization axis these atoms will experience a spin flip and the deceleration imparted by the cooling transition will be reduced. For example atoms transferring to the $m_{\mathrm{F}} =-9/2$ Zeeman sub-level will experience a maximum deceleration of only $\SI{3}{\metre \per \second\squared}$, because of the reduced Clebsch-Gordan coefficients. Without optical pumping these atoms will be lost. Luckily, the $\sigma^{+}$ pumping transition is an order of magnitude stronger, ensuring fast optical pumping of these atoms to the efficiently-slowed $m_{\mathrm{F}} = 9/2$ state.

As described above, optical pumping is important for the operation of the slower. We enhance it by using the dual frequency configuration for the slower beam, thereby improving the loading efficiency by a factor \StirringvsnSti~compared to when addressing only the cooling~transition. Note that only using the pumping~transition does not enable us to load any atoms in the MOT.

As atoms fall in the lower chamber they can fall out of resonance and be lost if they are not efficiently slowed, see Fig.~\ref{fig:Slower_Stage_Phase_Space_Plot}(a) black dashed line. To increase the available force, we ramp the detuning of both transitions addressed by the slower beam using an asymmetric sawtooth, as in Fig.~\ref{fig:SWAP_principle}. Operating the slower beam in this manner can yield a force up to twice the typical radiation pressure force \cite{Snigirev2019SWAPSrMOT}. Here, because atoms are always counter-propagating toward a single beam, ramps with positive or negative slope yield the same flux. Using an asymmetric sawtooth frequency sweep instead of a symmetric one serves only to suppress stimulated emission into the slower beam. Using the asymmetric frequency sweep we enhanced our loading efficiency by a factor of $\sim 1.5$ compare to using only the symmetric sweeps.
Here the sweep time has been independently optimized for each sweep configuration and we have used the laser detunings and powers shown in Table~\ref{table:CaptureMOT_parameters} (YMOT beam).

\section{Steady-state narrow-line MOT}
\label{Sec:MOT}

Atoms that are successfully slowed accumulate in a narrow-line 3D MOT operating on the \RedMotTransition~transition. In this section we describe how this MOT is set up and characterize its performance. In particular, we will focus on the comparison of the flux and lifetime between MOTs of \Sr{88} (bosonic strontium) and \Sr{87} (fermionic strontium).

The MOT confinement is provided by five beams, four in the horizontal plane and one propagating upward. The downward restoring force is provided by gravity. The upward propagating beam is also used as a slower beam, see Sec.~\ref{Sec:Slower}. The properties of these beams are given in Table~\ref{table:CaptureMOT_parameters}, while the field gradients used for the MOT are \fieldGradients~along the $\{x, y, z\}$ axes.

\begin{table}[tb]
 \begin{center}
 \caption{Properties of the MOT beams shown in Fig.~\ref{fig:Setup_schematic}. For each beam, we give the power of the cooling $||$ pumping beam, $1/e^{2}$ beam diameter $D$, and laser detuning sweep parameters: start detuning $\Delta_{1}$, stop detuning $\Delta_{2}$, and repetition rate $1/\Delta t$. Cooling and pumping beams have the same polarization, the same frequency modulation, and are combined within the same optical fibers before being sent to the atoms.}
 \label{table:CaptureMOT_parameters}
 \resizebox{\columnwidth}{!}{
 \begin{tabular}{lcccc}
 \hline \hline \noalign{\smallskip}
	 Beam & Power $\left( \si{\milli \watt}\right)$ & D $ \left( \si{\milli\meter} \right)$ & $\Delta_1 / 2 \pi; \Delta_2 / 2 \pi \left(\si{\mega\hertz} \right)$ & $1/\Delta t \left( \si{\kilo\hertz} \right)$ \\
 \noalign{\smallskip} \hline \noalign{\smallskip}
 	 YMOT & $~~9.2 ~||~6.6$ & $68$ & $-0.94;-3.49$ & $40$\\
	 XMOT1 & $~0.66~||~0.3$ & $47$ & $-0.65;-2.19$ & $15$\\
	 XMOT2 & $~ 0.66~||~0.3$ & $47$ & $-0.65;-2.19$ & $15$\\
	 ZMOT1 & $0.1~||~0$ & $8$ & $-0.82;-1.24$ & $17$\\
	 ZMOT2 & $0.1~||~0$ & $8$ & $-0.82;-1.24$ & $17$\\
 \hline \hline
 \end{tabular}
 }
 \end{center}
\end{table}

\begin{figure}[tb]
\includegraphics[width=0.98\columnwidth]{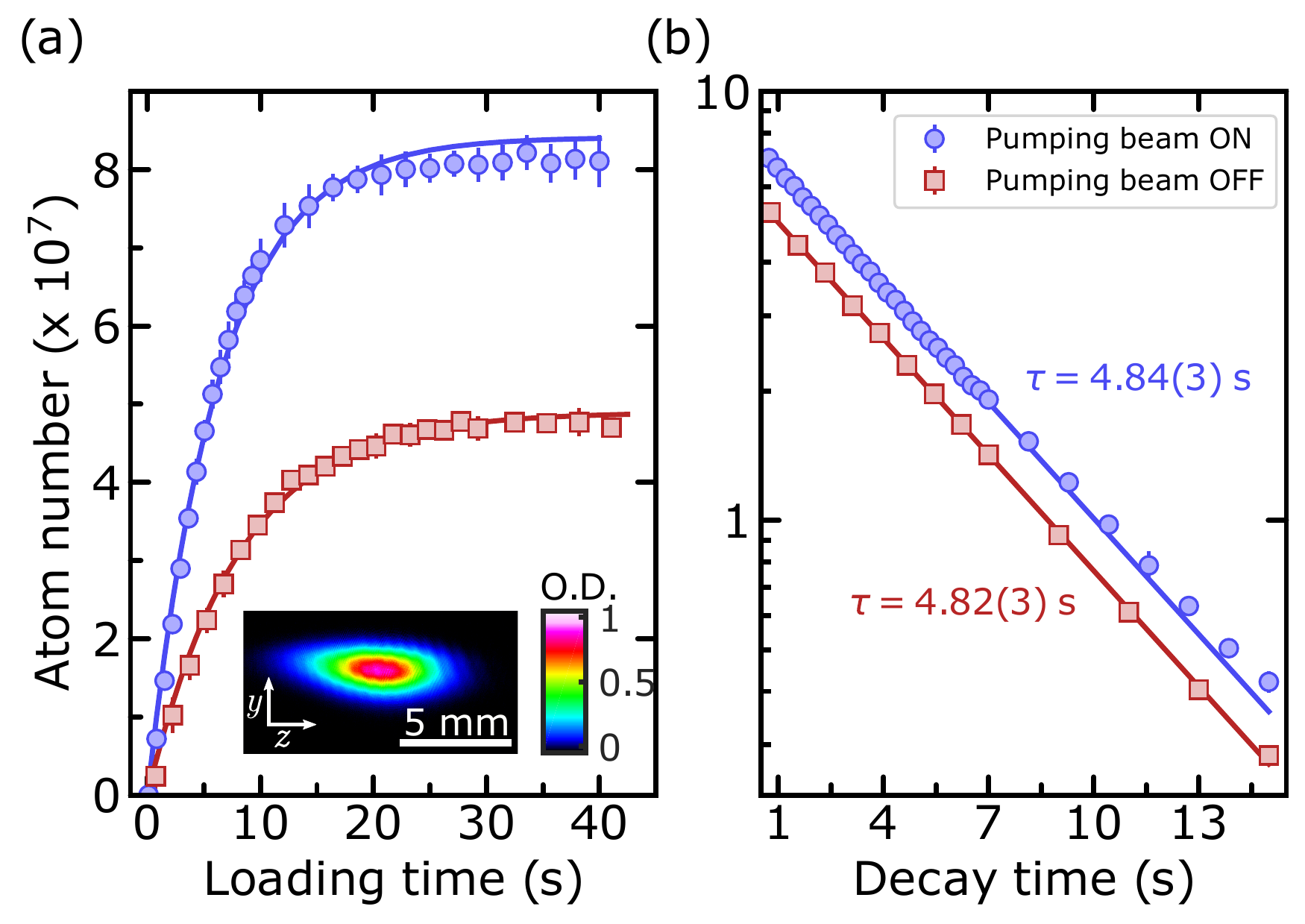}
\caption{\label{fig:MOT_Loading_Rate}  \Sr{87} MOT loading (a) and decay (b) with the pumping beams on and off. The solid lines are fits to the data using the functions (a) $N(t) = N_{\infty}(1-e^{-t/\tau_{\mathrm{load}}})$ and (b) $N(t) = N_{0}e^{-t/\tau_{\mathrm{life}}}$. When switching the pumping beams on, the flux into the MOT is enhanced by \StirringvsnSti~while the lifetime remains the same. In (a) at $t=0$ we switch the laser beams on. In (b), we load the MOT for $\SI{15}{\second}$, then switch off the 2D MOT in the top chamber at $t=0$. Error bars give the standard deviation from averaging $2-5$ data points. The inset in (a) shows a typical absorption image of the steady-state MOT, for $\SI{40}{\second}$ loading time in presence of the pumping beam.}
\end{figure}

In order to characterize the performance of this MOT, we measure the flux, temperature, steady-state atom number, and lifetime. We then compare the MOT for fermionic strontium with one for bosons produced with the same apparatus. The results are summarized in Table~\ref{table:MOT_values_comparison}. In order to characterize the MOT we take absorption images using the \BlueMotTransition~transition, see inset of Fig.~\ref{fig:MOT_Loading_Rate}.

To record the loading of the MOT, we repeat such measurements for different loading times after switching the laser beams on, see Fig.~\ref{fig:MOT_Loading_Rate}(a). As the flux of atoms entering the lower chamber is constant, we fit these data with the function $N(t) = N_{\infty}(1-e^{-t/\tau_{\mathrm{load}}})$, where $N_{\infty}$ is the steady-state atom number, and $\tau_{\mathrm{load}}$ is the $1/e$ loading time. We obtain from the fit $\tau_{\mathrm{load}} = $ \MOTLoadingTime~and $N_{\infty} = $ \MOTAtomNumber. We extract the flux of atoms $\Phi$ being loaded into the MOT as $\Phi = N_{\infty}/\tau_{\mathrm{load}} =$ \MOTflux. We also measure the $1/e$ lifetime by reaching steady state then turning off the 2D MOT in the upper chamber.

We measure the temperature of the steady-state MOT by time of flight and obtain \MOTYtemp~along the $y$ axis and \MOTZtemp~along the $z$ axis. As we use a single-axis imaging system, we are unable to measure the sample's size and temperature in the $x$ axis, and we assume it to be the same as in the other horizontal ($z$ axis) in order to calculate the average MOT temperature and peak density. However it must be noted that, because we tune the laser powers and frequencies independently for each axis, this approximation is rough.

\begin{table}[tb]
\begin{center}
 \caption{Characteristics of the steady-state MOT, for both \Sr{87} and \Sr{88} isotopes (see also description in the main text). The steady-state MOT extents $\sigma_{y,z}$ are Gaussian widths extracted by fitting a Gaussian function to the images of the MOT. All uncertainties are given as the standard deviation from the fitted data.}
 \label{table:MOT_values_comparison}
 \resizebox{\columnwidth}{!}{
 \begin{tabular}{lcc} 
 \hline \hline \noalign{\smallskip}
 & \Sr{87} MOT & \Sr{88} MOT\\
 \hline \noalign{\smallskip}
 Flux $\Phi$ ($\SI{}{atoms \per\second}$) & \MOTflux[] & \BosonicMOTflux[] \\
 $T_z$ ($\SI{}{\micro\kelvin}$) & \MOTZtemp[] & \BosonicMOTZtemp[] \\
 $T_y$ ($\SI{}{\micro\kelvin}$) & \MOTYtemp[] & \BosonicMOTYtemp[] \\ 
 $T_{\mathrm{average}}$ ($\SI{}{ \micro\kelvin}$) & \MOTAvgtemp[] & \BosonicMOTAvgtemp[] \\ 
 Width $\sigma_{y}$ ($\SI{}{\milli\meter}$) & \MotWidhty[] & \BosonicMOTWidhty[] \\ 
 Width $\sigma_{z}$ ($\SI{}{\milli\meter}$) & \MotWidhtz[] & \BosonicMOTWidhtz[] \\ 
 Atom number $N_{\infty}$ & \MOTAtomNumber[] & \BosonicMOTAtomNumber[] \\ 
 Peak density ($\SI{}{atoms\per\centi\meter\cubic}$) & \MOTDensity[] & \BosonicMOTDensity[] \\ 
 $1/e$ loading time $\tau_{\mathrm{load}}$ ($\SI{}{\second}$) & \MOTLoadingTime[] & \BosonicMOTLoadingTime[] \\
 $1/e$ lifetime $\tau_{\mathrm{life}}$ ($\SI{}{\second}$) & \MOTLifetime[] & \BosonicMOTlifetime[] \\
 \noalign{\smallskip} \hline \hline
 \end{tabular}
 }
\end{center}
\end{table}

We compare these data to an equivalent \Sr{88} bosonic MOT, using similar sweep parameters and magnetic fields. For \Sr{88} the collimating 2D molasses in the upper chamber uses the same detuning ramps and similar powers as the ones reported in our previous work \cite{Bennetts2017HighPSDMOT}. We follow the procedure described above to characterize this MOT and summarize the results in Table~\ref{table:MOT_values_comparison}. For bosons, the flux is \BosonicMOTflux, about \BosonNssoverfermionNss~times higher than for fermions. This difference can partly be explained by the 12 times higher natural abundance of \Sr{88} compared to \Sr{87} \cite{Laeter2003PureandappliedChemistry}. The remaining difference can be explained by the previously mentioned imperfections of the 2D molasses and Zeeman slower, owing to the more complex internal structure of \Sr{87}.

The similar lifetimes of \Sr{88} and \Sr{87} MOTs indicates that the two MOTs have similar losses. A well-known peculiarity of \Sr{87} narrow-line MOTs is that they have a loss mechanism not present in narrow-line bosonic MOTs, if no countermeasures are taken. This loss mechanism arises due to the difference in Land\'e factors between the ${^1\mathrm{S}_0} \, {\mathrm{F}=9/2}$ ground state and the ${^3\mathrm{P}_1} \, {\mathrm{F'}=11/2}$ excited state. Indeed, atoms within such MOTs can be optically pumped into states that are not trapped if they travel to the opposite side of the quadrupole magnetic field center. Typically these losses are avoided by adding a second set of laser beams addressing the pumping~transition. This transition generates ``stirring", a fast randomization of the population over all the $m_{\mathrm{F}}$ states \cite{Mukaiyama2003PRLRecoilLimitedLaser}. We apply this solution by using the pumping light described in Sec.~\ref{Sec:Slower} with the parameters reported in Table~\ref{table:CaptureMOT_parameters}. Fig.~\ref{fig:MOT_Loading_Rate}(b) compares the MOT's decay with and without pumping beams. Surprisingly, we found that the presence of the pumping light does not affect the \Sr{87} MOT lifetime. While the flux increases by a factor of \StirringvsnSti~the lifetime remains the same. 

This behavior can be explained by the small gradients of the MOT quadrupole field and large detuning of the MOT beams, combined with the sample's low temperature. Atoms slowed by the vertical beam and captured in the MOT are addressed at a distance $y_0 = \hbar \Delta_{1}/(g_{\mathrm{F'}} m_{\mathrm{F'}} \mu_{\mathrm{B}} \, dB/dy)$ from the MOT quadrupole magnetic field center, where $\mu_{\mathrm{B}}$ is the Bohr magneton. For $m_{\mathrm{F}} = +9/2$, $|y_0| \approx \SI{14} {\milli \meter}$ below the center, which is much bigger than the vertical Gaussian width $\sigma_y$ of the MOT sample. This large distance relative to the MOT size implies that atoms always lie well below the quadrupole field center. Here they are constantly addressed by the vertical beam and optically pumped mainly into the $m_{\mathrm{F}} = +9/2$ state. Therefore the losses observed in other \Sr{87} MOTs that are operated with less detuning and without stirring beam, are not present here. The convenient polarization mechanism at play here, resulting from weak optical forces, gravity and the large detuning of the MOT beams, has already been observed in MOTs of other elements such as dysprosium and erbium \cite{Ferlaino2012PRAFrisch,DreonJournalOfPhysicsB2017,AikawaPhysRevLett}. We note that, even if present, the pumping light is unlikely to have a strong effect on the MOT because it is strongly detuned at its location, see Fig.~\ref{fig:Slower_Stage_Phase_Space_Plot}. The role of the pumping light is to increase the atom flux $\Phi$ from the slower stage.

\section{Discussion and conclusion}
\label{Sec:Discussion}

Let us now compare the performance of our \Sr{87} MOT with the requirements for a superradiant active optical clock operating on the $\si{\milli\hertz}$-wide clock transition \cite{Meiser2010SteadyStateSuperradiance, Bohnet2012SuperradiantLaserRb, Norcia2016SuperradianceStrontiumClock}. Such a clock is a bad-cavity laser, which uses as a gain medium a cloud of ultracold atoms within the mode of a high-finesse optical cavity. After being optically pumped to an excited state, these atoms can emit into the cavity mode, with a rate of emission enhanced via collective, superradiant emission. To sustain the emission of this laser, the excited population must be renewed. Atoms could be renewed by pumping them back into the excited state, which leads to unfavorable heating, or by bringing new excited state atoms into the superradiantly lasing ensemble. This translates into requirements for a stream of new incoming atoms needed for allowing the laser's continuous-wave operation. For a laser operating on the \clocktransition~transition of strontium, we estimate that, without optical pumping, the flux of new atoms must be superior to $10^6 \, \si{^{87}Sr \, atoms \per \second}$. Moreover, atoms must be brought into the cavity mode, which for strontium can be achieved by a conveyor belt. For MOT atoms to be efficiently loaded into a conveyor of reasonable depth, their temperature should be lower than $\sim \SI{10}{\micro \kelvin}$. Our steady-state fermionic MOT satisfies these requirements. A lower temperature or higher density can be achieved by using less light intensity within MOT beams, to the detriment of the atomic flux. A higher flux can be straightforwardly achieved by operating at a higher oven temperature and/or by using ``repumper" beams to plug losses on the \BlueMotTransition~cooling cycle \cite{Dinneen1999Repump707679Sr}. 

To conclude, we have demonstrated a steady-state magneto-optical trap operating on the narrow intercombination line of strontium that works with the fermionic isotope \Sr{87}, despite the additional challenges arising from its hyperfine structure. Two key ingredients were necessary: (1) the use of 2D SWAP molasses to collimate the atomic beam before atoms pass from the upper vacuum chamber to the next; (2) a dual-frequency narrow-line Zeeman slower and MOT capable of slowing and capturing the incoming atoms. This MOT is well suited to provide atoms for the gain medium of a continuous-wave superradiant laser operating on strontium's  $\si{\milli\hertz}$-wide clock transition.

\begin{acknowledgments}
We thank the NWO for funding through Vici grant No. 680-47-619, and grant No. NWA.QUANTUMNANO.2019.002 (Quantum Inertial Navigation) and the European Research Council (ERC) for funding under Project No. 615117 QuantStro. This project has received funding from the European Union’s Horizon 2020 research and innovation program under grant agreement No 820404 (iqClock project). B.P. thanks the NWO for funding through Veni grant No. 680-47-438 and C.-C. C. thanks support from the MOE Technologies Incubation Scholarship from the Taiwan Ministry of Education. 

\end{acknowledgments}


%

\end{document}